\documentstyle[11pt,gh2001-asp,twoside,epsf]{article}

\begin{document}

\title{Simulations of dual morphology in spiral galaxies}
\author{Simon L. Berman}
\affil{Theoretical Physics, University of Oxford, 1 Keble Road, 
             Oxford, UK. simon@thphys.ox.ac.uk}

\begin{abstract}
Gas and stars in spiral galaxies are modelled with the {\sc dual}
code, using hydrodynamic and N-body techniques. The simulations reveal
morphological differences mirroring the dual morphologies seen in $B$
and $K'$ band observations of many spiral galaxies. In particular, the
gaseous images are more flocculent with lower pitch angles than the
stellar images, and the stellar arm-interarm contrast correlates with
the degree of morphological decoupling.
\end{abstract}

\section{Introduction}

There is much more to spiral galaxies than is apparent from simple
observations in optical light. Optical observations are dominated by
young stars and gas which may constitute only 5\% of a galaxy's
mass. The majority of the baryonic mass of a galaxy is only revealed
in infrared light. Despite the fact that young and old stars orbit in
the same potential, observations in the optical and near infrared can
reveal radical differences in morphology. Block \& Puerari (1999) show
that $B$ and $K'$ band images of the same galaxy can be completely
decoupled and that $K'$ band images are mostly one or two armed
whereas $B$ band images are frequently multi armed. Grosbol \& Patsis
(1998) find that grand design spirals are common in $K'$ band images,
but in the $B$ band, most galaxies are flocculent with tighter spiral
pitch angles of upto $7\deg$.

The {\sc dual} code, combining hydrodynamic and N-body techniques, has
been used to recreate these dual morphologies, and ask if predictions
can be made about intrinsic characteristics of spiral galaxies from
their morphologies. High density regions of hydrodynamic simulations
are assumed to reproduce the morphology of high luminosity regions in
$B$ band images, and N-body methods to trace the underlying mass
distribution of the galaxy seen in $K'$ band images.

\section{Galaxy models and the {\sc dual} code}

The models consist of freely evolving stellar and gaseous components,
and an analytic background dark halo. The motion of the stars is
deduced using a simple N-body solver, and the gas by solving the non
self-gravitating, isothermal Euler equations, using the FS2 algorithm
of van Albada, van Leer \& Roberts (1982). The units are the disk
scale length $R_d$ and the disk mass $M_d$ with $G$ set equal to
unity. Calculations take place on a 2D Cartesian grid of width 250
cells or 7.5 $R_d$, and with 2 $\times 10^5$ particles.

Stellar particles are drawn from the {\sc df} of Dehnen (1999) at
radii $R < 5$. Surface density and radial velocity dispersion profiles
are exponential, with scale lengths set to give a constant scale
height, equated to the N-body softening length. The slope of the
rotation curve at $4 R_d$ is set to zero and the halo mass and scale
length are fixed by choosing $f_d$, the disk mass fraction at $4
R_d$. Choosing the minimum value of the stability parameter $Q$ fixes
the magnitude of the radial velocity dispersion. Gas starts on
circular orbits at uniform surface density. The sound speed is 0.025
(10 km s$^{-1}$ if $R_d$ = 5 kpc and $M_d = 2 \times 10^{11} \, {\rm
M}_{\sun}$).

\section{Analysis and results}

To highlight the underlying spiral structures, the images are
subjected to a 2D Fourier transform using logarithmic spirals as basis
states. The pitch angle $\psi$ is related to the number of arms $m$
and the radial wavenumber $\alpha$ by $\tan \psi = - m / \alpha$.

Figure 1 is representative of many of the simulations, showing a grand
design spiral in both stars and gas. Its Fourier spectra corroborate
this: both the stars and gas are strongly peaked in the $m = 2$
spectrum. Since the gaseous arms are narrow, the gas also possess
significant power in the $m = 4$ and $m = 6$ spectra. The stellar
image of Figure 2 is also strongly two armed, verified by the
corresponding spectra. The gaseous image may seem a mess, but the
spectra shows a strong $m = 4$ structure with a similar, but smaller,
pitch angle to the stars. This is similar to the morphology of our
Galaxy deduced from observations by Drimmel \& Spergel (2001) and
simulations by Englmaier \& Gerhard (1999) and Fux (1999). Lastly, the
stellar structure of Figure 3 is also two armed, as seen from its
spectra. However, the gas image is now completely flocculent and its
spectrum is strongly peaked at all values of $m$. Many interfering
spiral waves prevent any single spiral dominating the image.

The flocculence of the images has been quantified by realizing that
grand design spirals have one strong peak in their spectrum and
flocculent spirals are multiply peaked, and hence more uniform. The
uniformity of a distribution is measured by its entropy, $S=-{\sum_i}
p_i \ln p_i$. Ignoring the $m$ = 4 and 6 spectra, the entropy is
converted into a flocculence 0 $\geq \mathcal{F} \geq$ 1 by dividing
by the maximum entropy (due to a flat distribution) $S_{\max} = \ln
N_{\max}$, where $N_{\max} = 1604$ is the total number of components
in the $m = 1,2,3$ and $5$ spectra.

\begin{figure}
\plottwo{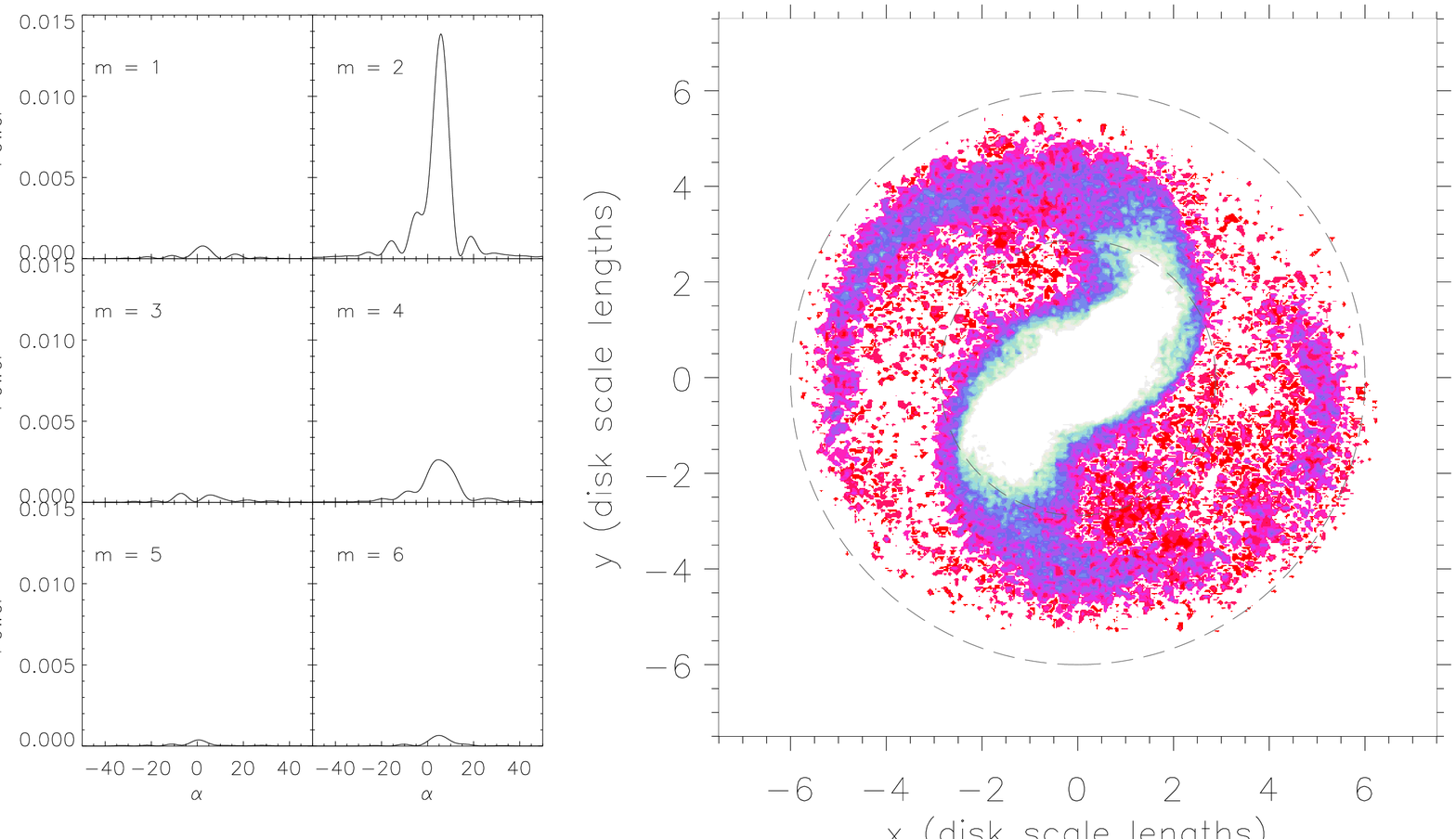}{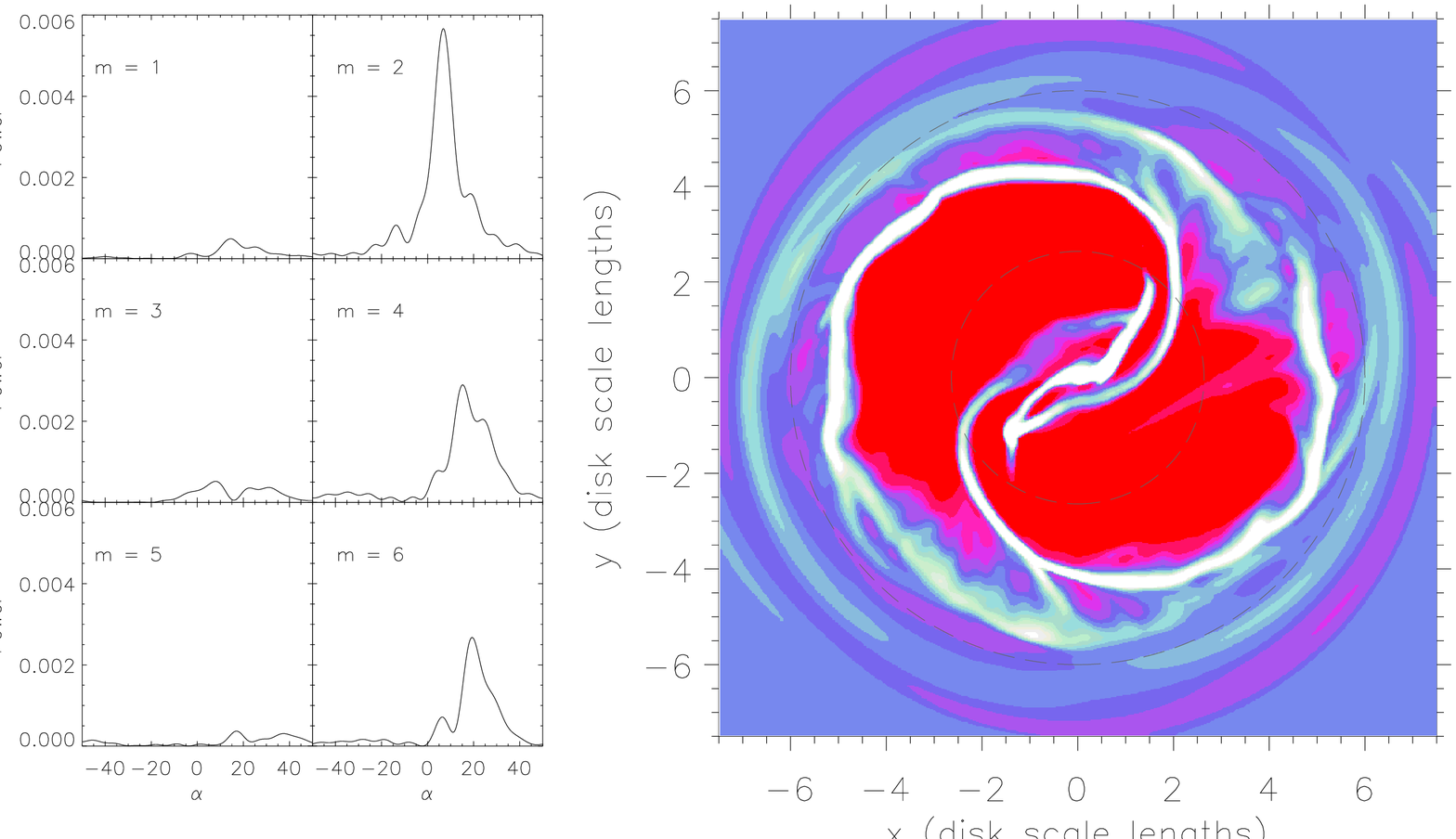}
\caption{Stellar and gaseous density fields and harmonic spectra for a grand design spiral. Dashed lines indicate minimum and maximum radii used in the Fourier analysis. Flocculence and pitch angles are ${\cal F}_{\rm sta} = 0.51$ and $\psi_{\rm sta} = 20.9^{\circ}$ for the stars and ${\cal F}_{\rm gas} = 0.53$ and $\psi_{\rm gas} = 16.5^{\circ}$ for the gas. Stellar arm-interarm contrast is ${\cal C} = 2.48$.}
\end{figure}

\begin{figure}
\plottwo{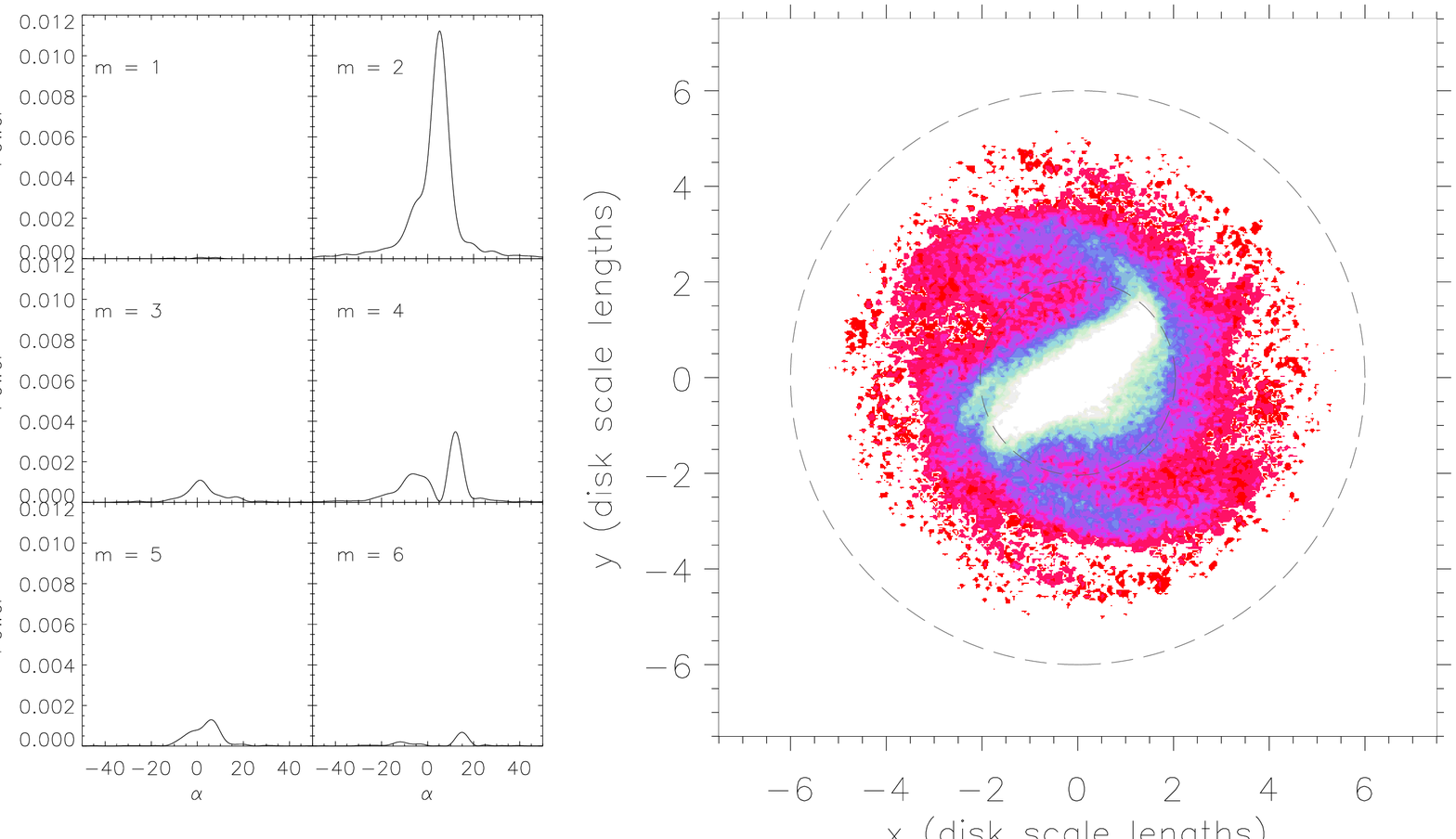}{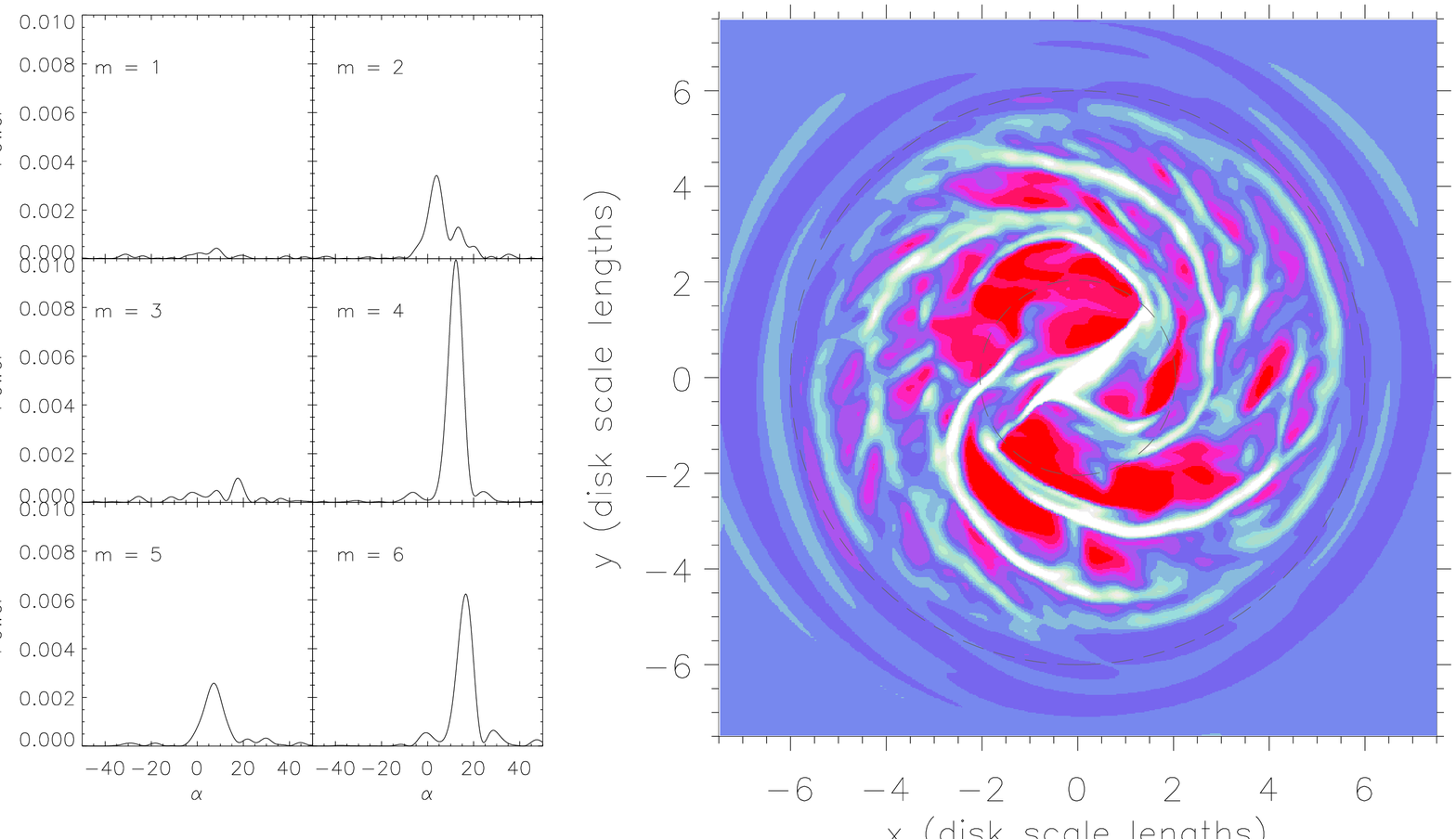}
\caption{A multi armed spiral. Flocculence and pitch angles are ${\cal F}_{\rm sta} = 0.53$, $\psi_{\rm sta} = 20.9^{\circ}$ (two arms) for the stars and ${\cal F}_{\rm gas} = 0.61$, $\psi_{\rm gas} = 17.7^{\circ}$ (four arms) for the gas. Stellar arm-interarm contrast is ${\cal C} = 1.69$.}
\end{figure} 

\begin{figure}
\plottwo{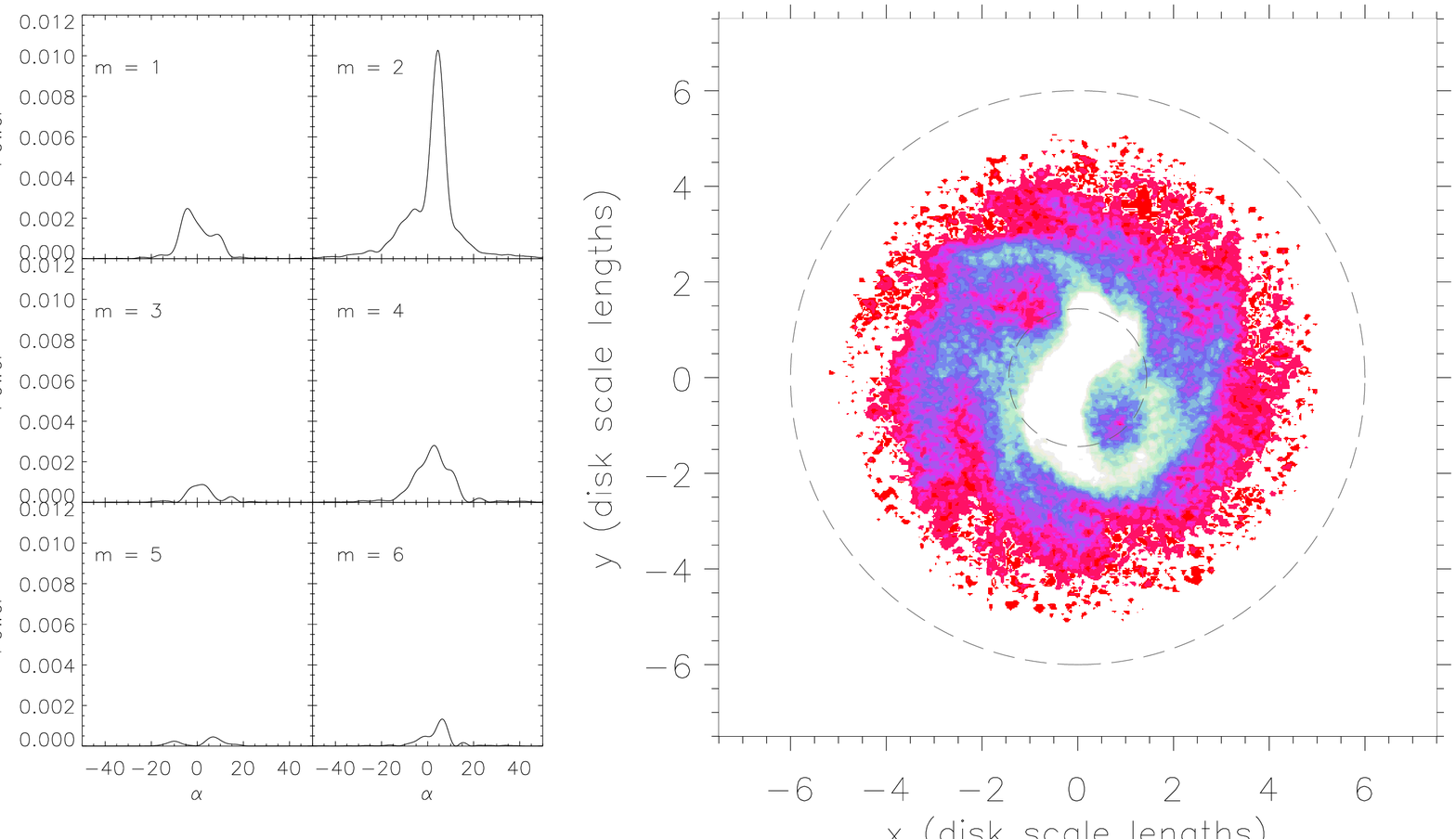}{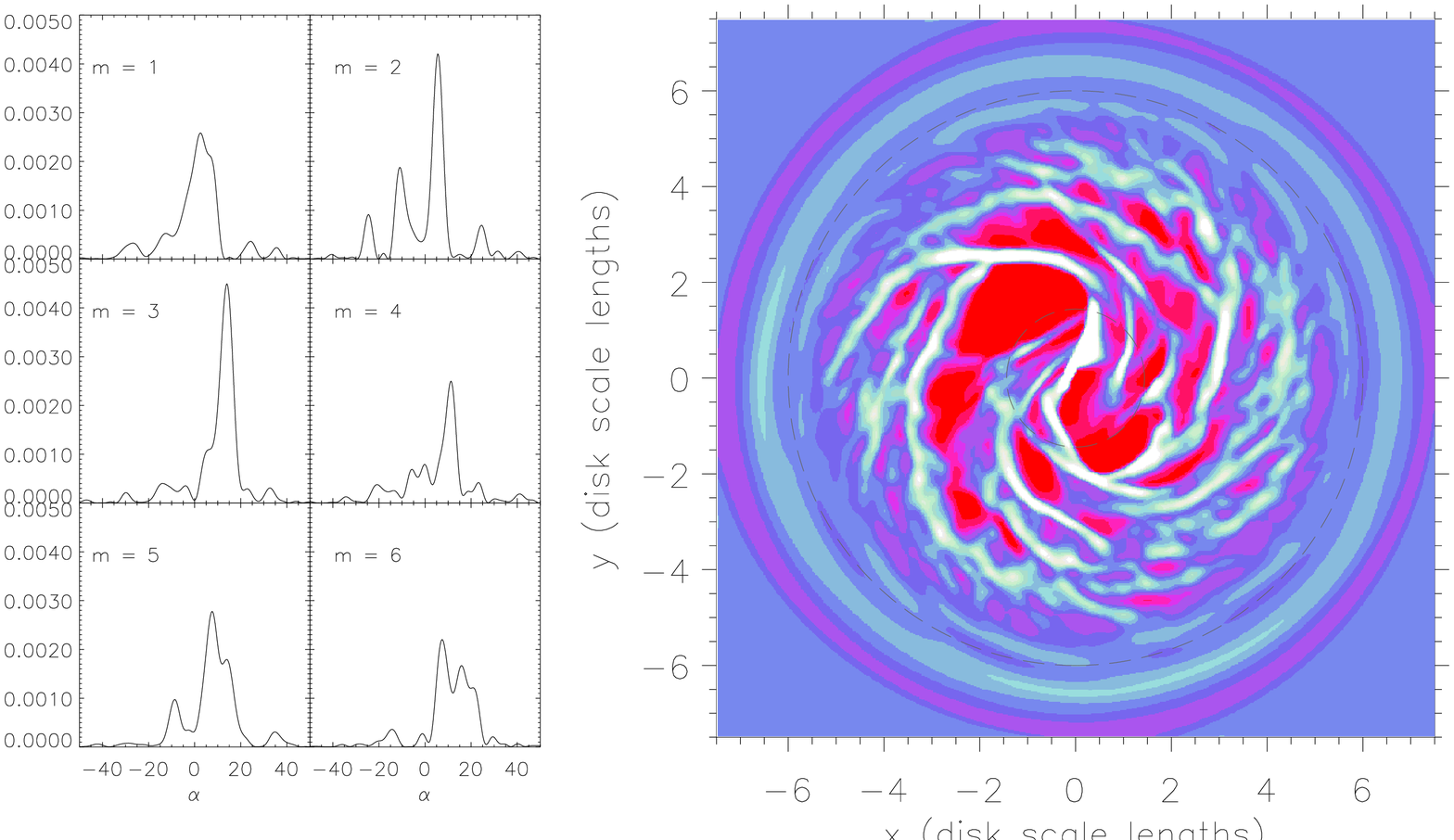}
\caption{A flocculent spiral. Flocculence and pitch angles are ${\cal F}_{\rm sta} = 0.47$, $\psi_{\rm sta} = 24.0^{\circ}$ (two arms) for the stars and ${\cal F}_{\rm gas} = 0.70$ for the gas. Stellar arm-interarm contrast is ${\cal C} = 1.58$.}
\end{figure} 

The figures show that dual morphologies seen in $B$ and $K'$ band
observations are recreated in simple simulations. Figure 4 shows that
the gaseous pitch angles are similar to or tighter than in stellar
images. It also shows that the stellar images are less flocculent than
gaseous ones. In fact, almost all of the stellar models have a smooth
two armed grand design structure, and the gaseous images exhibit many
extra ragged small scale structures that are not seen in the stars.

The arm-interarm contrast and the gaseous flocculence correlate with a
coefficient of $-0.77$ (see Figure 4). Since $K'$ band and stellar
images are generally not flocculent, the gas flocculence, and hence
the arm-interarm contrast, is a guide to the degree of decoupling
between stellar and gaseous morphologies. The stellar arm-interarm
contrast is probably also the cause of decoupling: a low contrast
enables the noise in the stellar component to form low amplitude
features in the gas and a flocculent structure. But with a high
contrast, the stellar spiral dominates, creating a grand design
gaseous spiral. Thornley (1996) observed 4 optically flocculent
galaxies in the $K'$ band and found evidence for grand design
structures with arm-interarm contrasts of just 1.1 to 1.4.

The mean value of $Q$ correlates with gaseous flocculence and stellar
arm-interarm contrast (coefficients of $-0.62$ and 0.81). This results
from stellar particles scattering off spiral arms, with higher
contrast leading to more scattering and hotter disks. However,
simulations evolve non-linearly, so it is not possible to predict the
initial $Q$ from the end state of a run. A simulation's history has a
large effect on the morphology at any one time. Seeds in the noise of
the stellar distribution allows small scale structure to develop. The
low sound speed and shocks in the gas amplifies these structures and
stops them dissapearing quickly.

\begin{figure}
\plotone{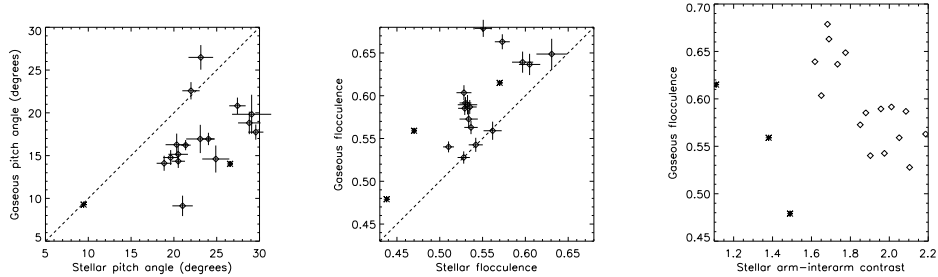}
\caption{Left two panels: Gaseous and stellar pitch angle $\psi$ and flocculence $\mathcal{F}$ for all simulations. Lines of equal pitch angle or flocculence are dashed. Right panel: Correlation of arm-interarm contrast $\mathcal{C}$ with gaseous flocculence $\mathcal{F}_{\mathrm{gas}}$. Asterisks indicate values for real galaxies.}
\end{figure} 

\section{Conclusion}

Simple simulations have reproduced some of the differences in spiral
galaxy morphology in the optical and infrared. These differences can
be explained by the dynamics of stars and gas, without invoking
interactions, star formation or dust obscuration. In particular, gas
images are more flocculent with lower pitch angles than stellar
images. Stellar arm-interarm contrast and $Q$ correlate inversely with
the degree of decoupling. The effect of gas self-gravity is not known,
but it is needed to model the full range of galaxies
self-consistently.


\begin{references}

\reference Block, D. L., \& Puerari, I. 1999, \aap, 342, 627

\reference Dehnen, W. 1999, \aj, 118, 1201

\reference Drimmel, R., \& Spergel, D. N. 2001, \apj, 556, 181

\reference Englmaier, P., \& Gerhard, O. 1999, \mnras, 304, 512

\reference Fux, R. 1999, \aap, 345, 787

\reference Grosbol, P. J., \& Patsis, P. A. 1998, \aap, 336, 840

\reference Thornley, M. D. 1996, \apj, 469, L45

\reference van Albada, G. D., van Leer, B., \& Roberts, W. W. 1982, \aap, 108, 76

\end{references}
\end{document}